\documentclass[10pt, a4paper]{article}
\usepackage{amsmath, amssymb}
\usepackage{graphicx, subcaption, authblk,setspace, indentfirst}
\usepackage[a4paper, total={7in, 9in}]{geometry}

\title{Two dimensional, electronic particle tracking in liquids with a graphene-based magnetic sensor array}
\author[1]{R. F. Neumann}
\affil[1]{IBM Research, Av. Pasteur 138 \& 146, Urca, Rio de Janeiro, 22290-240, Brazil}
\author[2]{M. Engel}
\author[1]{M. Steiner}
\affil[2]{IBM Research, Yorktown Heights, New York, 10598, United States}

\begin{document}
\maketitle
\onehalfspacing

\begin{abstract}
The investigation of liquid flow at the nanoscale is a key area of applied research with high relevance to Physics, Chemistry and Biology.
We introduce a method and a device that allows to spatially resolve liquid flow by integrating an array of graphene-based magnetic sensors used for tracking the movement of magnetic nanoparticles in the liquid under investigation.
With a novel device conception based on standard integration processes and experimentally verified material parameters, we simulate the performance of a single sensor pixel, as well as the whole array, for tracking magnetic nanoparticles.
The results demonstrate the ability (a) to detect individual nanoparticles in the liquid and (b) to reconstruct particle trajectories across the sensor array as a function of time, in what we call ``Magnetic Nanoparticle Velocimetry'' technique.
Being a non-optical detection method, potential applications include particle tracking and flow analysis in opaque media at sub-micron scales.
\end{abstract}

\section{Introduction}

The characterization of liquid behavior at the nanoscale, including wetting phenomena and flow, is a key objective for research in Physics, Chemistry, and Biology with potential technological applications across various industries. Liquids confined in microscale and nanoscale porous media exhibit different physical behaviors than those in larger dimensions, mainly due to the increased surface-to-volume ratio, leading to enhanced surface effects, and to the breakdown of the continuum framework for flow~\cite{eijkel2005nanofluidics, bocquet2010nanofluidics}. A method that allows to measure liquid flow with nanometer scale resolution would enable the analysis of liquid-solid interactions that are not typically captured in continuum flow models.

Characterizing flow in very small constrictions is typically achieved by tracking  individual particles immersed in the liquid under study using optical microscopy methods~\cite{yoda2007nano, park2004nano}. These methods require optically transparent samples, calling for an all-electronic detection method which could be integrated with non-transparent samples. Indeed, significant progress in the integration of electronic sensing functionalities for the characterization of fluids has been made~\cite{chen2014convex, lohse2015surface, nirmalraj2014nanoelectrical, leong2014dynamics, steiner2015nanowetting, jeon2005optically}. However, a sensor (array) that can be integrated with microfluidic or nanofluidic devices and able to track electronically the movement of a single nanoparticle $(< 100\,\textnormal{nm})$ has yet to be demonstrated. Most demonstrations so far lack the sensitivity to detect single nanoparticles, the ability to track moving nanoparticles, or have limited spatial resolution due to large active sensor areas $(\gtrsim 10 \mu\textnormal{m}^2)$.

Previous attempts at detecting small magnetic particles employing Si-~\cite{besse2002detection, mihajlovic2007magnetic}, InSb-~\cite{di2011detection}, and InAs/AlSb~\cite{mihajlovic2007submicrometer} Hall devices targeted the detection of magnetic microbeads composed by thousands of nanometer-sized iron-oxide particles dispersed in a polymer matrix.
Analogously, Al$_2$O$_3$-~\cite{shen2005situ} and MgO-based~\cite{shen2008detection, shen2008quantitative} Magnetic Tunnel Junctions have also been used to detect such magnetic microbeads.
Giant magnetoresistance spin valve sensors~\cite{li2006spin} or semiconductor-based Hall sensors~\cite{di2010single, manzin2012modelling} were used to detect compound nanobeads made of hundreds or thousands of sub-$20\,\textnormal{nm}$ particles, provided they were attached directly to the sensor.

Graphene, due to its favorable electronic properties, has been identified as candidate material for electronic flow sensing. Flow sensing based on single-cell detection was achieved employing a graphene transistor array integrated into a microfluidic channel~\cite{ang2011flow}. Moreover, all-electronic detection of the average flow speed in a microchannel was demonstrated by measuring the streaming potential of an electrolyte solution through the changes it caused in the electrical properties of a graphene transistor integrated into the microfluidic chip~\cite{he2012solution, newaz2012graphene}. Based on its potential to integrate within fluidic devices and to shrink lateral dimensions below the micrometer scale, a graphene-based Hall sensor~\cite{panchal2013epitaxial, panchal2012small} could be used to detect single magnetic nanoparticles immersed in a liquid in nanoscale constrictions. Furthermore, the integration of a two-dimensional sensor array within a fluidic channel device would provide an all-electronic means of tracking a particle's position through the simultaneous analysis and correlation of Hall-voltage changes in each individual sensor of the array.

\section{Results and Discussion}

We discuss in the following an integration framework and investigate numerically an all-electronic, device-based method to map liquid flow at the nanometer scale by tracking an individual magnetic nanoparticle immersed in the liquid under study. To that end, we introduce the conception of an array of graphene-based Hall-effect sensors which is integrated with a fluid channel device based on realistic manufacturing steps. We perform a numerical analysis based on experimentally confirmed material parameters that reveals quantitatively the sensitivity of  individual nanoparticle detection in the liquid as a function of the particle position. A simultaneous analysis of Hall voltages from $N \times M$ graphene-based Hall detectors forming a sensor array allows to reconstruct the two-dimensional particle trajectory within the channel device.

\subsection{Device implementation and principle of operation}

\begin{figure}[ht]
\centering
\includegraphics[width=\columnwidth]{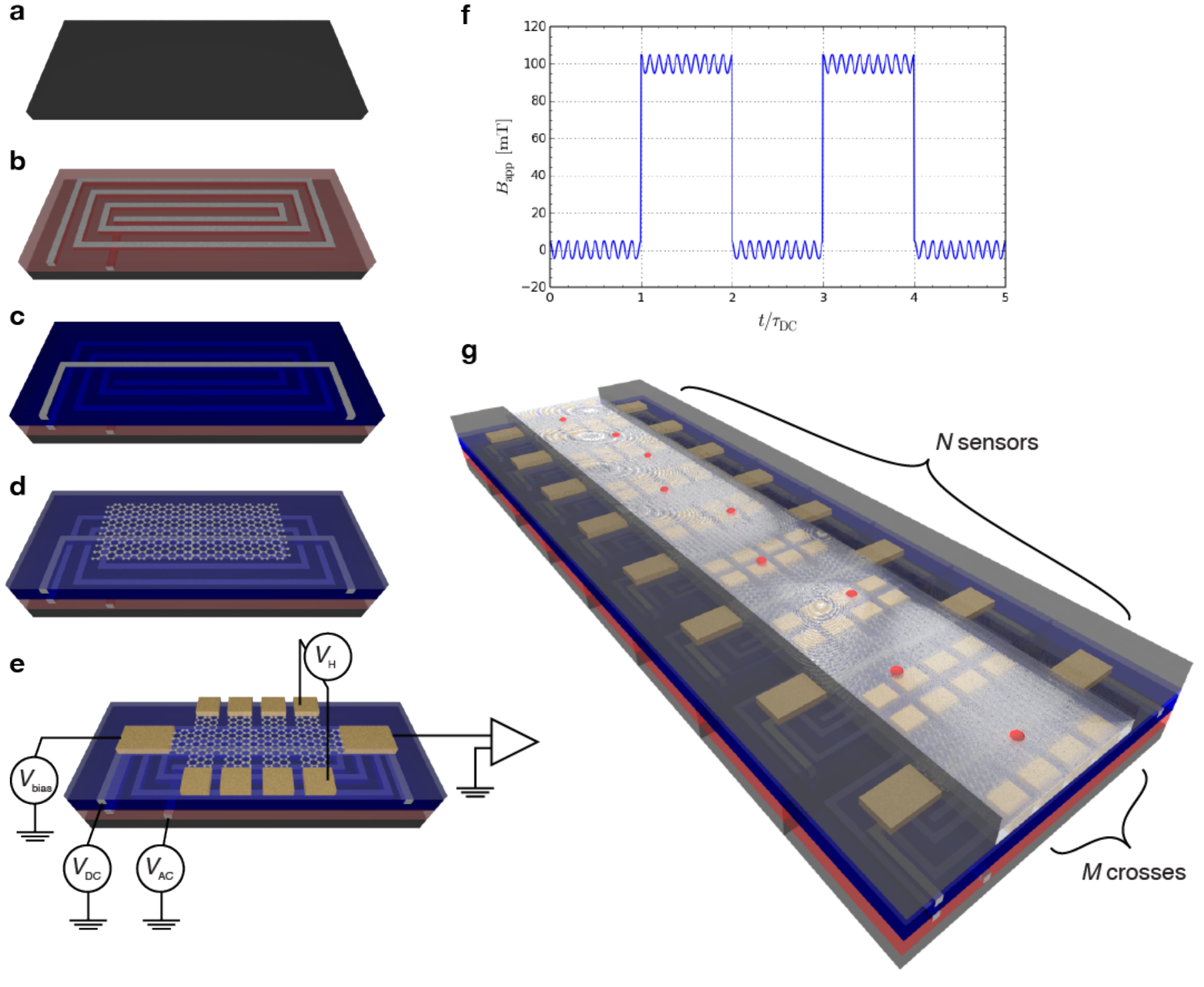}
\caption{(a)-(e) 3D illustration of the process flow for monolithic integration of (b) co-planar spiral inductor for DC magnetic field generation, (c) metal line for AC magnetic field generation, (d) graphene transfer, (e) Hall device definition and electrical interfacing. Also shown are the electrical connections and working principle.
(f) Applied field $B_\textnormal{app}$ as a function of time, depicting $B_\textnormal{AC} = 5\,\textnormal{mT}$ and $B_\textnormal{DC} = 100\,\textnormal{mT}$ components.
While the AC component is always on, the DC part is turned on and off at intervals $\tau_\textnormal{DC}$.
(g) Platform comprising $N$ sensors with $M$ crosses each, combined with a microfluidic channel that guides magnetic nanoparticles.}
\label{fig:device}
\end{figure}

Based on basic integration steps and standard manufacturing principles, we introduce an approach to implement an integrated, two dimensional nanoscale flow sensor. The sensing scheme presented here requires on-demand generation of DC and AC magnetic fields to facilitate static and dynamic magnetization in conjunction with a Hall sensor device. In Figure~\ref{fig:device}(a)-(e) we illustrate the monolithic integration of the different functional layers.

We start with a planar substrate that in principle can be solid or flexible like, for example, silicon or polyimide, as in Figure~\ref{fig:device}(a). In Figure~\ref{fig:device}(b) we show the first functional layer which is a co-planar spiral inductor that is used to generate a DC magnetic field. First a dielectric layer is deposited followed by patterning and pattern transfer by means of dry etching. Next metal is evaporated in the patterned trenches followed a planarization step. This process sequence is repeated with a via etch in between to create a underpass metal line that provides connection to the inner contact of the spiral~\cite{Reyes:1995bv}. Optionally, a ferromagnetic core can be included into the spiral to increase the DC magnetic field~\cite{Ahn:1993br}. In Figure~\ref{fig:device}(c) we illustrate the second functional layer that is a metal line to produce a AC magnetic field. Fabrication is similar to the previous layer except that the process has only to be done once. To insulate the second functional layer from the Hall sensor device another dielectric layer is deposited followed by transfer of graphene on top of the dielectric surface by a suitable process~\cite{Li:2009bv}, as illustrated in Figure~\ref{fig:device}(d). Next the Hall bar structure is patterned and etched into the graphene by for example oxygen plasma. In a final step metal contacts to the graphene are defined followed by metallization and a metal lift-off process, as shown in Figure~\ref{fig:device}(e).

Figure~\ref{fig:device}(e) also shows the working principle of the sensor. The Hall device is voltage biased ($V_\textnormal{bias}$) while the appropriate DC and AC magnetic fields are generated by independently applying voltages $V_\textnormal{DC}$ and $V_\textnormal{AC}$, respectively. The interplay between the $B_\textnormal{DC}$ and $B_\textnormal{AC}$ components of the applied field $B_\textnormal{app}$ is depicted in Figure~\ref{fig:device}(f).
A small AC field $B_\textnormal{AC} \sim 10^0\,\textnormal{mT}$ is continuously applied. It induces an AC magnetization on the nanoparticle, that should be in-phase with the applied field provided that the period of oscillation $\tau_\textnormal{AC}$ is larger than the nanoparticle's relaxation time~\cite{neel1949theorie}. A stronger DC field $B_\textnormal{DC} \sim 10^2\,\textnormal{mT}$ is switched on and off at regular intervals $\tau_\textnormal{DC} \gg \tau_\textnormal{AC}$. The DC field (partially) saturates the nanoparticle, therefore decreasing its susceptibility and the amplitude of the AC component of the magnetization. The variation of the Hall voltage $V_\textnormal{H}$ AC amplitude between the ``on'' and ``off'' states of the DC field is directly affected by the presence of a nanoparticle nearby. In Figure~\ref{fig:device}(g) we illustrate the proposed sensing platform in combination with a fluidic channel on top that guides magnetic nanoparticles across the Hall sensor devices.

The $V_\textnormal{H}(t)$ signal measured by the sensor is modified in presence of a nanoparticle and, therefore, may be used to extract its location $\vec{r}_\textnormal{0}(t)$ within the channel.
The integrated device conception illustrated in Figure~\ref{fig:device}(g) enables tracking a magnetic nanoparticle's position with respect to the embedded sensor array.
This device geometry, with $N$ sensors and $M$ measuring pads per sensor, permits the collection of $N \times M$ independent Hall voltage signals $V_\textnormal{H}(t)$ -- a task which might be performed in parallel or sequentially.
The matrix $\left[ V_\textnormal{H} (t) \right]_{n,m}$ contains the information for localizing nanoparticles and for estimating their flow velocity inside the channel.
This matrix can be thought of as a two-dimensional pixel array, with each pixel providing a certain voltage signal that reflects the presence of a nearby nanoparticle.
As a nanoparticle crosses the array, the pixels indicate the nanoparticle's presence through changes in their respective Hall voltages. By analyzing and correlating the signals of a set of ``pixels'' it is, hence, possible to determine a particle's position, velocity, and trajectory with respect to the sensor array.

In view of applications, we note that when considering a superparamagnetic nanoparticle solution, a liquid such as water should be used so that diamagnetic effects can be neglected.
Also, the nanoparticle concentration should be adjusted in order to avoid agglomeration; see Appendix. In addition, surface coating and functionalization within the fluid channel may be considered as well~\cite{lu2007magnetic}.

\subsection{Simulated device operation and sensitivity assessment}

\begin{figure}[ht]
\centering
\includegraphics[width=\columnwidth]{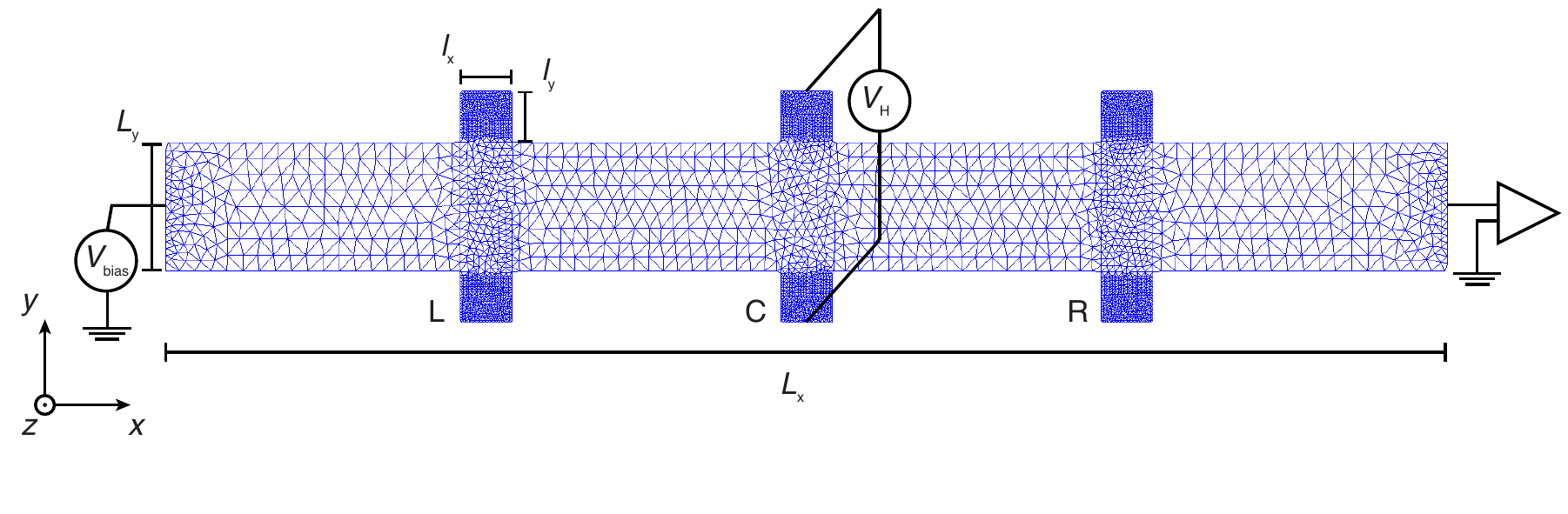}
\caption{Sketch of the computational mesh and sensor geometry with indicated voltages.
The $V_\textnormal{H}$ Hall voltage signal due to the total magnetic field (including the spatially inhomogeneous nanoparticle field) is calculated by solving the electronic transport equations using a finite-element approach.}
\label{fig:sensor-sketch}
\end{figure}

In the following, we discuss simulations of the device operation obtained by numerically solving the charge continuity equation in an ohmic medium (Equation~\ref{eq:pde}) under the influence of the spatially inhomogeneous magnetic field (Equation~\ref{eq:BNP}) provided by a nanoparticle.
The following results assume a graphene sensor with a ``triple cross'' geometry containing three measuring pads (L / C / R), as the one in Figure~\ref{fig:sensor-sketch}, with the following dimensions: $L_{\textnormal{x}} = 1\,\mu\textnormal{m}$, $L_{\textnormal{y}} = 100\,\textnormal{nm}$ and $\ell_{\textnormal{x}} = \ell_{\textnormal{y}} = 40\,\textnormal{nm}$.
We use realistic electronic parameters for modeling the graphene sheet~\cite{geim2007rise, chen2008intrinsic, akturk2008electron}, i.e. $\rho = 10^{16}\,\textnormal{m}^{-2}$ and $\mu = 1.5\,\textnormal{m}^2 \textnormal{V}^{-1} \textnormal{s}^{-1}$.
A DC voltage $V_\textnormal{0} = 100\,\textnormal{mV}$ is applied along $x$ by an external source in order to impose a current flow.
Whenever a magnetic field acts upon the current, a transverse Hall voltage $V_\textnormal{H}$ emerges in one or more of the measuring pads.
Unless stated otherwise, the applied magnetic fields were $B_\textnormal{AC} = 5\,\textnormal{mT}$ and $B_\textnormal{DC} = 100\,\textnormal{mT}$.
For more information on the choice of magnetic fields, see Appendix.
All calculations were performed at $T = 300\,\textnormal{K}$ but no significant changes were observed to within $50\,\textnormal{K}$ of that value.
The nanoparticle saturation magnetization was chosen as $M_\textnormal{S} = 380\,\textnormal{kA/m}$ for it is a typical value for widely-available iron-oxide nanomaterials~\cite{dunlop2001rock, cullity2011introduction}.

The change in the AC component of the Hall voltage when the DC field is applied is the chosen figure-of-merit, for it is highly sensitive to the nanoparticle's position.
We represent this figure-of-merit as $\delta V_\textnormal{H}^{\gamma}$ (Equation~\ref{eq:delta_VH}), where $\gamma$ controls whether there is $(\gamma = 1)$ or not $(\gamma = 0)$ a nanoparticle nearby.
The relationship between Equation~\ref{eq:delta_VH} and its experimental realization can be understood in terms of the applied field in Figure~\ref{fig:device}(f).
While $\delta V_\textnormal{H}^\textnormal{0}$ is the baseline signal and needs to be measured only once (before the nanoparticle solution is inserted), the detection signal $\delta V_\textnormal{H}^\textnormal{1}$ needs to be measured continuously over time.
The minuend on the left corresponds to the measurement of the AC Hall voltage during the times when the DC field is ``off'', while the subtrahend on the right corresponds to the measurement performed when the DC field is ``on''.
By continuously measuring the AC amplitude on both the ``on'' and ``off'' states of the DC field, by means of integrated circuitry, one can monitor the presence of a nearby magnetic nanoparticle with time resolution of $2 \tau_\textnormal{DC}$, simply by comparing the detection signal to the baseline.

Using the previously mentioned values for the parameters, the baseline level was calculated to be at $\delta V_\textnormal{H}^{0} = 0.56\,\mu\textnormal{V}$.
The intensity of the $\delta V_\textnormal{H}^{1}$ Hall voltage depends on the size $D$ and position $\vec{r}_\textnormal{0}$ of the nanoparticle.
In general, typical values of the figure-of-merit $\delta V_\textnormal{H}^{\gamma}$ range from $10^{0}$ to $10^{2}\,\mu\textnormal{V}$, so that the relative change $\Omega_m = \left( \delta V_\textnormal{H}^{1} / \delta V_\textnormal{H}^{0} \right)\!\big|_{m}$ in the Hall voltage AC amplitude caused by a nearby nanoparticle at a given measuring pad $m$ can exceed a 100-fold increase.
For more details on the calculation, see Appendix.

\subsection{Single-pixel performance and nanoparticle detection}

\begin{figure}[ht]
\centering
\begin{subfigure}[h]{0.48\textwidth}
    \caption{\hfill~}
    \includegraphics[width=\columnwidth]{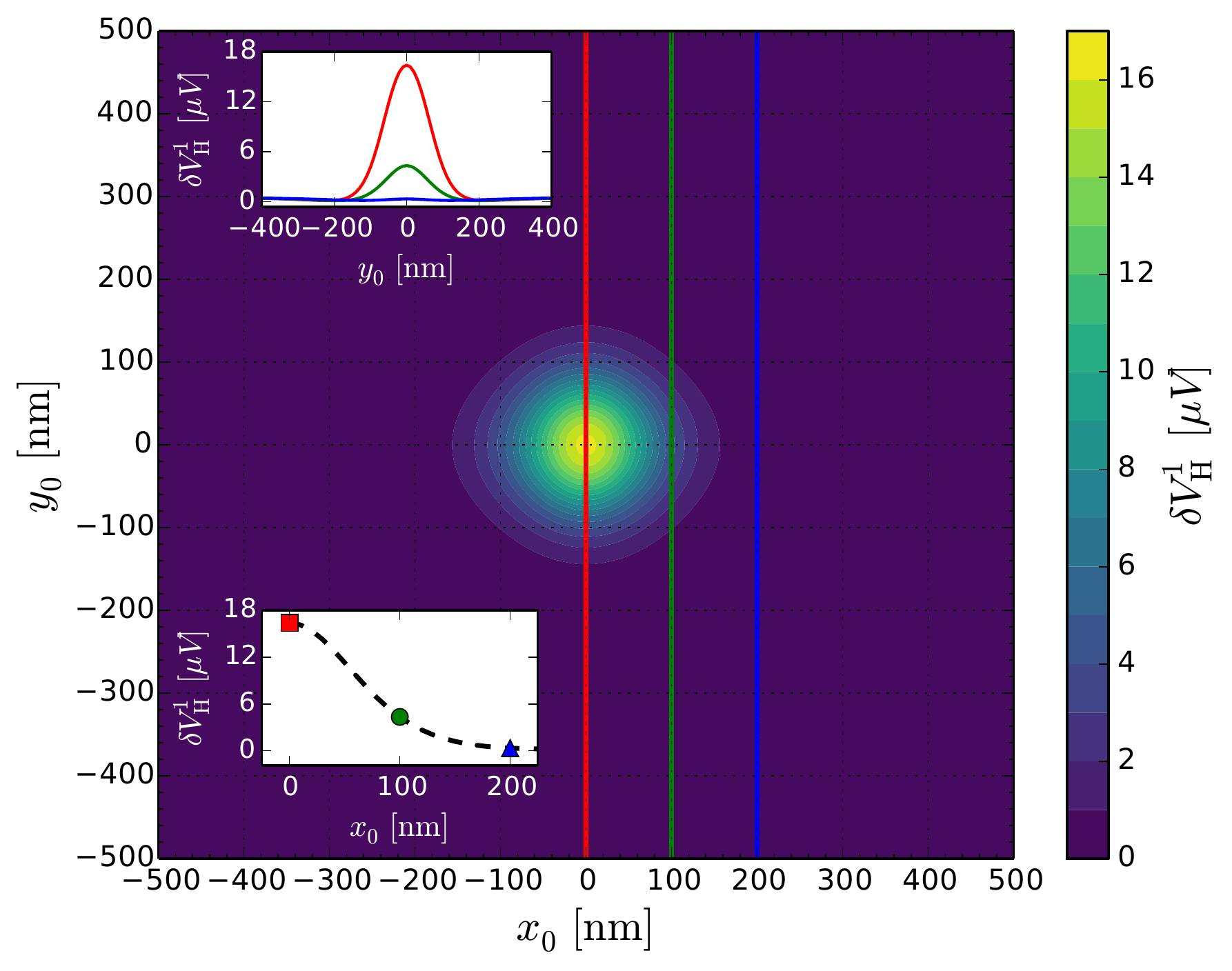}
    \label{fig:XY}
\end{subfigure}
\begin{subfigure}[h]{0.48\textwidth}
    \caption{\hfill~}
    \includegraphics[width=\columnwidth]{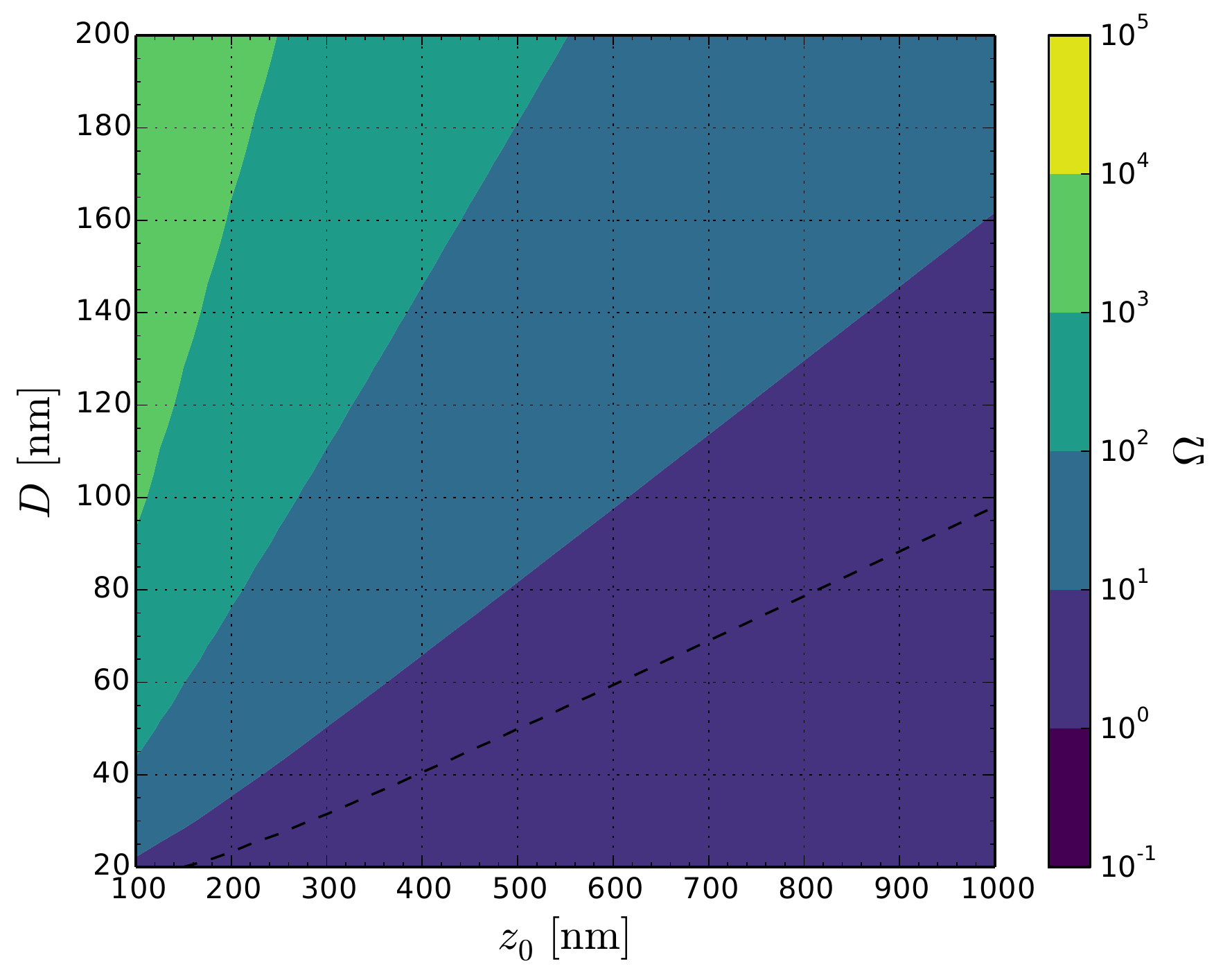}
    \label{fig:Omega-Dz}
\end{subfigure}
\caption{Single-pixel detection sensitivity demonstrated by numerical calculations.
(\subref{fig:XY}) Detection signal $\delta V_\textnormal{H}^{1}$ for a $D = 30\,\textnormal{nm}$ nanoparticle at several positions $(x_\textnormal{0}, y_\textnormal{0})$ in the $z_\textnormal{0} = 100\,\textnormal{nm}$ plane.
The upper inset graph shows the detection signal plotted along the color-coded lines in the main graph.
The lower inset graph shows the value of each curve on the upper inset graph for $y_\textnormal{0} = 0\,\textnormal{nm}$.
(\subref{fig:Omega-Dz}) Two-dimensional plot of the relative change $\Omega$ as a function of the nanoparticle diameter $D$ and vertical distance $z_\textnormal{0}$ for nanoparticles directly above the central measuring pad $(x_\textnormal{0} = 0\,\textnormal{nm}, y_\textnormal{0} = 0\,\textnormal{nm})$.
The dashed line represents $\Omega = 3$.}
\label{fig:single_pixel}
\end{figure}

In order to simulate the detection of a magnetic nanoparticle as it moves across a single sensor, the Hall detection signals at the central measuring pad were calculated for several nanoparticle positions $\vec{r}_\textnormal{0}$.
Figure~\ref{fig:single_pixel}(\subref{fig:XY}) shows the detection signals $\delta V_\textnormal{H}^{1}$ obtained for a $D = 30\,\textnormal{nm}$ nanoparticle at positions $(x_\textnormal{0}, y_\textnormal{0})$ in the $z_\textnormal{0} = 100\,\textnormal{nm}$ plane.
The maximum value corresponds roughly to a 30-fold increase with respect to the baseline level.
One can see in the graph that the effect of the magnetic field is strongly localized within $200\,\textnormal{nm}$ around the point on the sensor surface below it.
Numerical analysis of Equation~\ref{eq:BNP} confirmed by calculations for different $z_\textnormal{0}$ indicate that the half-width of the peak is about $2 z_\textnormal{0}$.
The upper inset in Figure~\ref{fig:single_pixel}(\subref{fig:XY}) shows that, as the nanoparticle moves along $y$ on the $x_\textnormal{0} = 0\,\textnormal{nm}$ line (in red), a bell-shaped voltage peak is detected as a function of time by the central sensor, reaching a maximum $\Omega_\textnormal{C} \sim 30$-fold increase with respect to the baseline level $\delta V_\textnormal{H}^{0} = 0.56\,\mu\textnormal{V}$.
When the $x$-coordinate of the nanoparticle moves away from the central measuring pad, the voltage signal decreases drastically, as can be seen in the lower inset graph of Figure~\ref{fig:single_pixel}(\subref{fig:XY}).

According to these results, and using Figure~\ref{fig:sensor-sketch} as geometrical reference, nanoparticles flowing along $y$ in the middle of the channel $(x_\textnormal{0} \approx 0\,\textnormal{nm})$ generate voltage signals mainly on measuring pad ``C'', while nanoparticles flowing closer to the left (right) channel wall generate voltage signals mainly on the ``L'' (``R'') measuring pad.
Therefore, the resolution for nanoparticle detection in $x$ ($y$) is related to the distance $\Delta x$ ($\Delta y$) between the pixels.
In other words, the higher the pixel density, the better the spatial resolution.

For greater vertical distances, larger nanoparticles are required to maintain an $\Omega$-value sufficient for single particle detection.
The dependence of $\Omega$ on the nanoparticle diameter $D$ and vertical distance $z_\textnormal{0}$ is shown in Figure~\ref{fig:single_pixel}(\subref{fig:Omega-Dz}), for a nanoparticle located directly above the central measuring pad.
For example, at $z_\textnormal{0} = 500\,\textnormal{nm}$ $(1000\,\textnormal{nm})$, a $D = 50\,\textnormal{nm}$ $(100\,\textnormal{nm})$ nanoparticle is required to achieve a moderate sensitivity of $\Omega \approx 3$, as illustrated by the dashed line. The nanoparticle size should, therefore, be selected depending on the channel height and the desired sensitivity threshold of the sensor. In essence, a sub-$100\,\textnormal{nm}$ magnetic particle will be detectable within fluid channels having heights of up to $1\,\mu \textnormal{m}$.
For a more detailed study of the size and distance dependence, see Appendix.

\subsection{Multi-pixel nanoparticle tracking and trajectory reconstruction}

\begin{figure}[ht]
\centering
\begin{subfigure}[h]{0.3\textwidth}
    \caption{\hfill~}
    \includegraphics[width=\columnwidth]{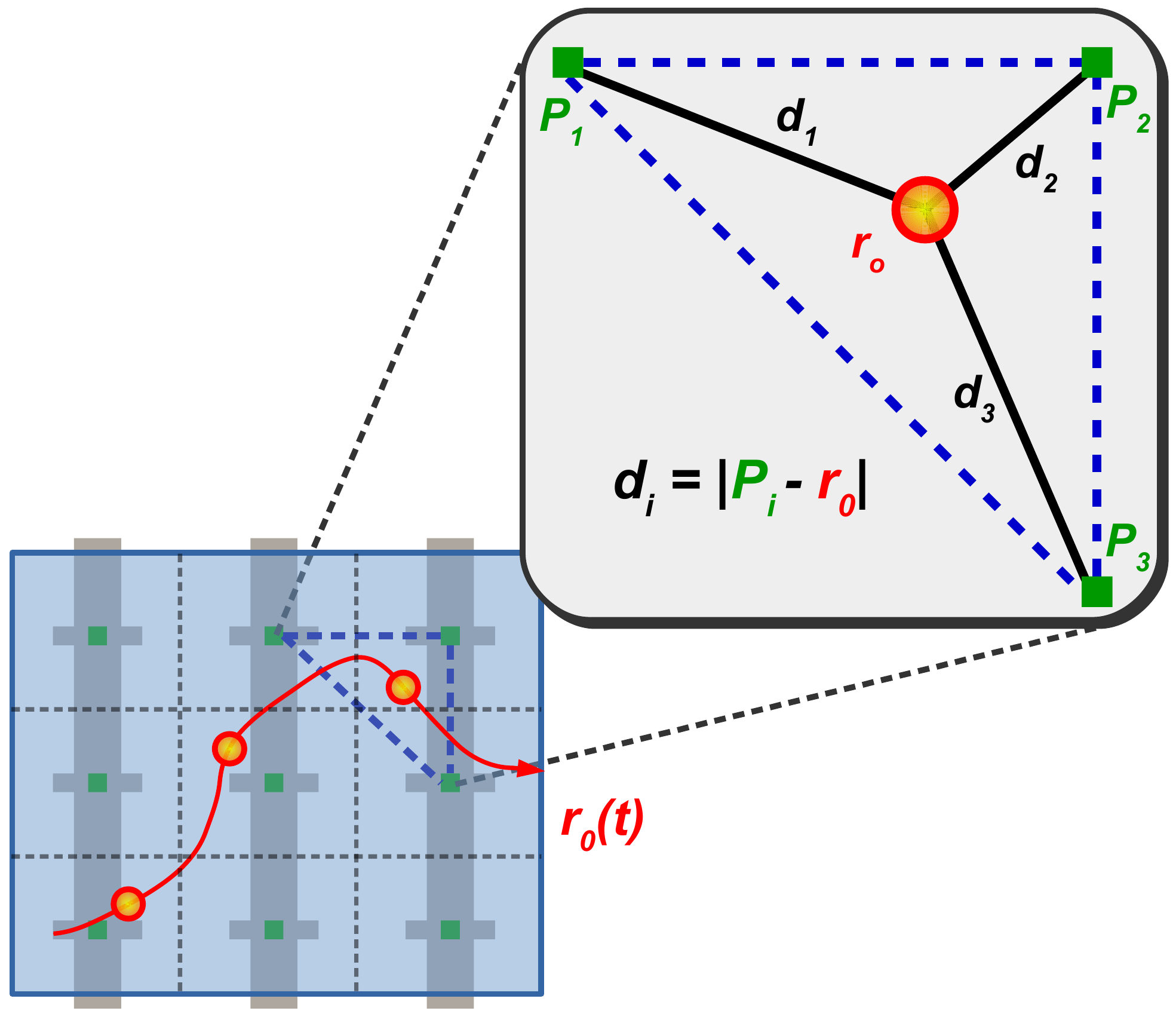}
    \label{fig:matrix}
\end{subfigure}
\begin{subfigure}[h]{0.32\textwidth}
    \caption{\hfill~}
    \includegraphics[width=\columnwidth]{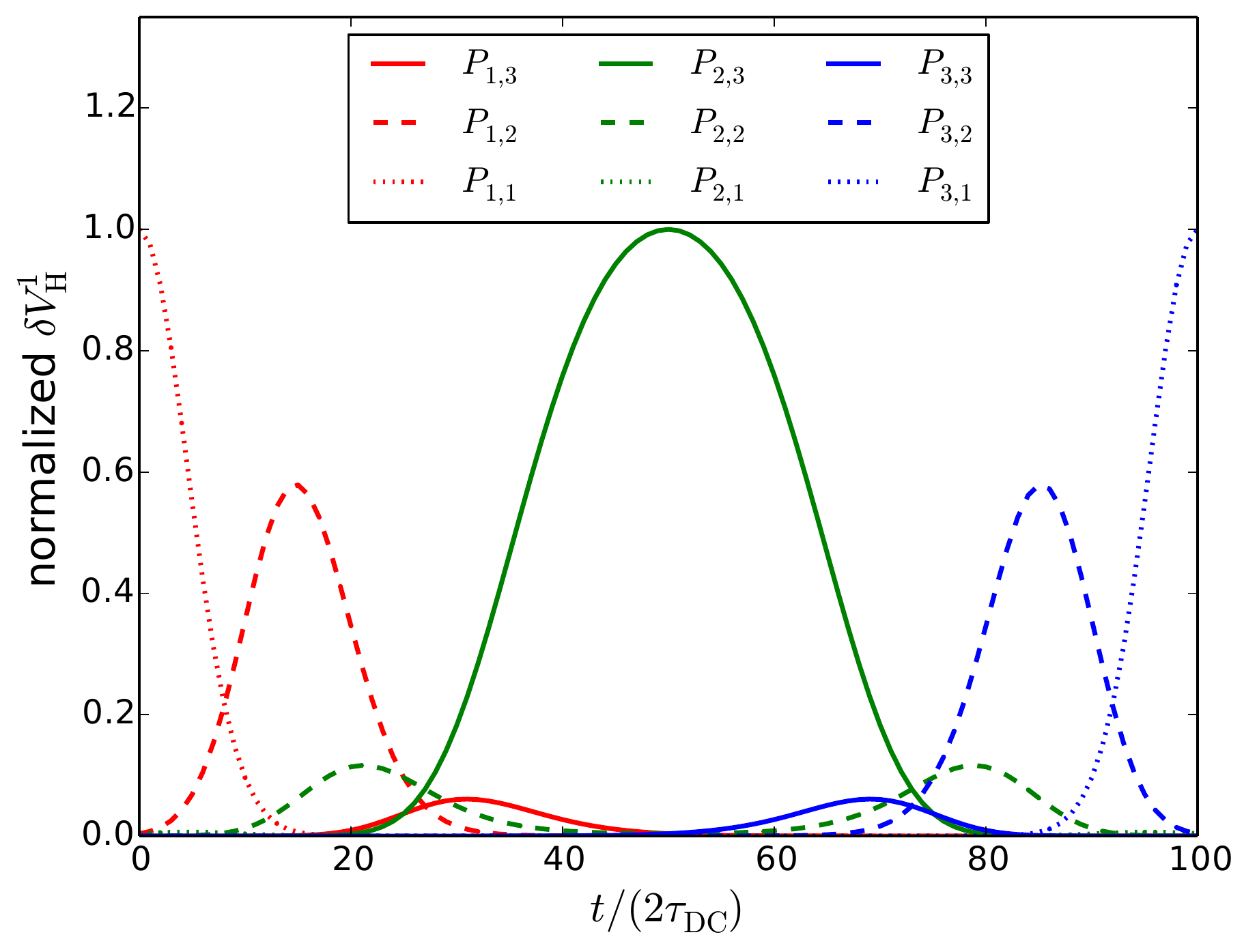}
    \label{fig:voltages}
\end{subfigure}
\begin{subfigure}[h]{0.32\textwidth}
    \caption{\hfill~}
    \includegraphics[width=\columnwidth]{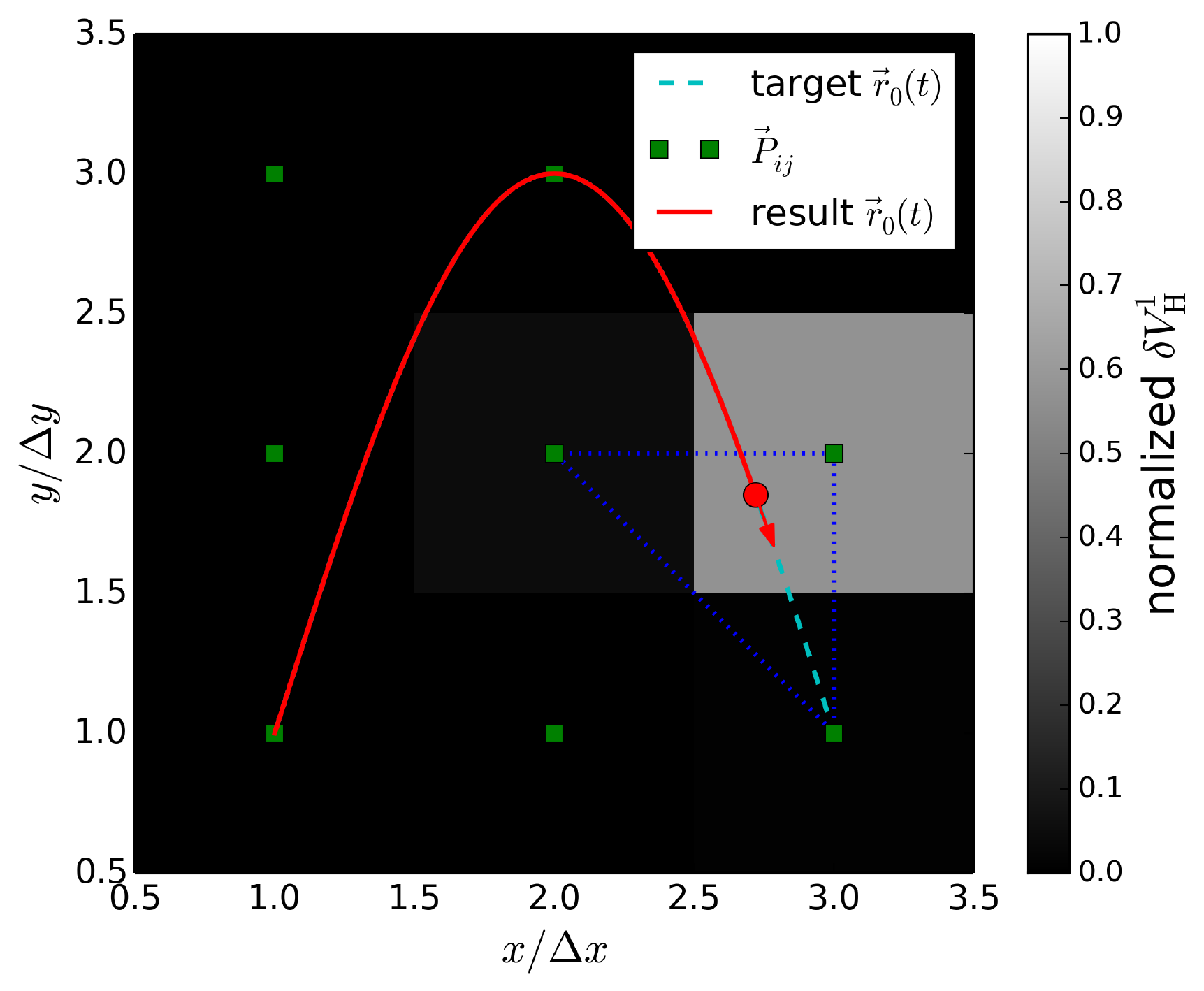}
    \label{fig:trajectory}
\end{subfigure}
\caption{(\subref{fig:matrix}) Sketch of device implementation with $3 \times 3$ Hall crosses, integrated into a fluidic channel (light blue) where a nanoparticle (red) moves with an arbitrary trajectory $\vec{r}_\textnormal{0}(t)$.
The dashed grid overlay represents the analogy between the sensing units and a pixel array (green squares).
The inset illustrates the trilateration algorithm used to extract particle positions from voltage signals.
Panel (\subref{fig:voltages}) shows the set of nanoparticle-induced voltage signals used in our computational experiment, while (\subref{fig:trajectory}) shows the reconstructed trajectory and velocity vector (red) at $t / (2 \tau_\textnormal{DC}) = 86$.}
\label{fig:tracking}
\end{figure}

Having established the detection capabilities of an individual, graphene-based Hall sensor, we will now investigate the detection capabilities of a two-dimensional sensor array. In order to illustrate the trajectory reconstruction methodology, we generated synthetic data simulating the movement of a nanoparticle across a two-dimensional sensor array with $N \times M$ Hall crosses that collect a matrix of $\delta V_\textnormal{H}^{1}$ voltage signals as a function of time, as shown in Figure~\ref{fig:tracking}(\subref{fig:matrix}).
In this particular example, we use a $3 \times 3$ pixel array located at points $\vec{P}_{i j} = \left( i \Delta x,  j \Delta y, 0 \right)$ with $(i, j) \in \{1, 2, 3\}$.
The response function of pixel $(i, j)$ was modeled as a Gaussian centered at $\vec{P}_{i j}$ with $\sigma_{x} = \sigma_{y} = 0.3$.
The particle trajectory in this example was $\vec{r}_\textnormal{0} (t) = \left( 1 + \frac{t}{50}, 1 + 2 \sin \frac{\pi t}{100}, 0 \right)$ for $t \in [0, 100]$, which generated the voltage signals depicted in Figure~\ref{fig:tracking}(\subref{fig:voltages}).
At every time step, the distances $d_{i j} = |\vec{P}_{i j} - \vec{r}_\textnormal{0}| = \mathtt{Gauss}^{-1} \left( [ \delta V_\textnormal{H}^{1} ]_{i j} \right)$ to the pixel centers were calculated and the three closest pixels used to define a triangle $\{P_1, P_2, P_3\}$, as in Figures~\ref{fig:tracking}(\subref{fig:matrix}) and~\ref{fig:tracking}(\subref{fig:trajectory}).
Using a standard trilateration algorithm~\cite{sturgess1995trilateration}, the nanoparticle position $\vec{r}_\textnormal{0}$ was calculated at each time step as a function of the distances $\{d_1, d_2, d_3\}$ and sensor coordinates $\{P_1, P_2, P_3\}$, reconstructing the full trajectory.
The nanoparticle velocity $\frac{\mathrm{d}}{\mathrm{d} t}\vec{r}_\textnormal{0}(t)$ was calculated using finite differences.
The final results in Figure~\ref{fig:tracking}(\subref{fig:trajectory}) demonstrate agreement between simulated and reconstructed trajectories.
In general, an experimental characterization of the voltage-distance response function is required to calibrate the distances given the measured voltage signals.
In order to illustrate the reconstruction of the nanoparticle trajectory from the voltage signals we provide a video in the Appendix.

We would like to point out that there are general rules-of-thumb that may be considered while choosing the several length and time constants involved in the method, for they impose limits on the trajectories and velocities that can be measured.
Each measurement of $\delta V_\textnormal{H}^{\gamma}$ requires the definition of an integration window $\tau_\textnormal{int}$ for the lock-in amplifier.
We suggest choosing $\tau_\textnormal{int} = \tau_\textnormal{DC} \gtrsim 10 \tau_\textnormal{AC}$, so that every plateau of $B_\textnormal{DC}$ encompasses several oscillations of $B_\textnormal{AC}$ and allows for the integration of the signal by the lock-in amplifier.
By synchronizing the lock-in amplifier to $B_\textnormal{DC}$ and setting $\tau_\textnormal{int} = \tau_\textnormal{DC}$, every measurement corresponds to either the minuend or subtrahend of Equation~\ref{eq:delta_VH}, so that a new value for $\delta V_\textnormal{H}^{1}$ is obtained every $2 \tau_\textnormal{DC}$.
The frequency of data collection is the largest limitation to the fidelity of reconstructed trajectories and velocities.

In terms of applications, the graphene-based sensor array could be integrated with ``lab-on-chip'' immunoassay devices as magnetic nanoparticles are widely used for separation, biosensing and sampling~\cite{kim2005magnetic, nikitin2007new, do2008polymer, ng2012digital}.
In addition, the array could be integrated to measure fluid flow properties in micro- or nanochannels made of opaque materials, such as ``reservoir-on-chip'' micromodels using geo-materials~\cite{gunda2011reservoir, porter2015geo, song2014chip}.
The potentially high spatial resolution of the array allows the investigation of liquid-solid interactions, such as wettability and slip phenomena, which could unveil novel aspects of the fluid's behaviour at the nanometer scale. Finally, an extension of the graphene-based sensor conception to other 2D materials is needed in order to take advantage of the tunable electronic properties of novel semiconductor materials.

\section{Conclusions}
In summary, we have investigated a graphene-based Hall-effect sensor array that is integrated within a fluid channel device for characterizing liquid flow at the nanoscale. We have introduced an implementation concept based on CMOS-compatible manufacturing steps and, based on experimentally verified materials parameter, we have performed numerical simulations for quantifying integrated device operation and detection sensitivity.
Based on our analysis results, the system is capable of (1) performing nanoparticle detection by means of a single sensor pixel and (2) to record nanoparticle trajectories based on the simultaneous analysis of multiple voltage signals in case a multi-pixel sensor array is used.
A sensitivity analysis indicates that sub-$100\,\textnormal{nm}$ magnetic particles are detectable within fluid channels having heights of up to $1\,\mu \textnormal{m}$. The device conception and the ``Magnetic Nanoparticle Velocimetry'' method has potential as a flow sensing technique for the characterization of liquid behavior at the nanoscale, including wetting phenomena and flow, with potential technological applications in industrial and applied research.

\section*{Computational method}

The Hall device was modelled as a two-dimensional conducting sheet at $z = 0$ with carrier density $\rho$ and carrier mobility $\mu$, representing a graphene sheet~\cite{geim2007rise}.
The geometry of the sheet is composed of a rectangular section with dimensions $(L_\textnormal{x}, L_\textnormal{y})$ and a set of $M$ measuring pads or ``crosses'' with dimensions $(\ell_\textnormal{x}, \ell_\textnormal{y})$ equally spaced along the rectangular section by a distance $\Delta x = L_{\textnormal{x}} / (M + 1)$.
The sensor is oriented perpendicular to the fluidic channel so that nanoparticles flow along $y$ and the channel width is comparable to $L_{\textnormal{x}}$.

The electrical transport in the sensor is described by its space-dependent conductivity tensor $\bar{\sigma}(x,y)$, with components $\sigma_\textnormal{xx} = \sigma_\textnormal{yy} = \sigma_\textnormal{0} / \left[ 1 + (\mu B)^2 \right]$ and $\sigma_\textnormal{xy} = -\sigma_\textnormal{yx} = \mu B \sigma_\textnormal{xx}$, where $\sigma_0 = \rho e \mu$ is the zero-field conductivity and $B(x,y)$ is the inhomogeneous magnetic field perpendicular to the sensor surface.
Combining the continuity equation  $\nabla \cdot \vec{J} = 0$ with Ohm's law $\vec{J} = \bar{\sigma} \vec{E}$ leads to the PDE that provides the electrostatic potential $\phi(x,y)$
\begin{equation}
\nabla \cdot \left[ \bar{\sigma}(x,y) \nabla \phi(x,y) \right] = 0
\label{eq:pde}
\end{equation}
under Dirichlet boundary conditions to enforce $\phi(L_\textnormal{x}/2, y)  - \phi(-L_\textnormal{x}/2, y) = V_\textnormal{0}$.
The Hall voltage $V_\textnormal{H}$ for a given magnetic field is calculated by taking the difference of the average potential at opposite sides of a given measuring pad.
Due to the inhomogeneous nature of Equation~\ref{eq:pde}, its solutions can only be obtained numerically~\cite{ibrahim1998diffusive, liu1998effect, bending1997hall}, a task for which we used the PDE solver \texttt{FreeFem++}~\cite{hecht2012new, jouault2008finite}.
The computational mesh employed is similar to that of Figure~\ref{fig:sensor-sketch}, only twice as dense, containing 1000 points along the contour.
A self-adaptive mesh was also tested, but results turned out to be more numerically stable and reproducible with an homogeneous mesh.

A superparamagnetic nanoparticle located at position $\vec{r}_\textnormal{0} = (x_\textnormal{0}, y_\textnormal{0}, z_\textnormal{0})$ inside the liquid, is subject to a homogeneous applied field $B_\textnormal{app}$, with DC and AC components such that $B_\textnormal{DC} \gg B_\textnormal{AC}$.
The magnetization of such nanoparticle at temperature $T$ is given by a Langevin function $M(B_\textnormal{app}, T) = M_\textnormal{S} \mathcal{L} \left( M_\textnormal{S} V B_\textnormal{app}/k_\textnormal{B} T \right)$,
where $V = (\pi/6) D^3$ is the volume of the nanoparticle \cite{neel1949theorie, bean1959superparamagnetism}.
The magnetic dipole field created by the nanoparticle's magnetic moment at (and perpendicular to) the $z = 0$ plane is given by
\begin{equation}
B_\textnormal{NP}(x,y) = \frac{\mu_0}{4 \pi} M V \left[ \frac{3 z_0^2 - r(x,y)^2}{r(x,y)^5} \right] \, ,
\label{eq:BNP}
\end{equation}
where $r(x,y) = \sqrt{(x - x_0)^2 + (y - y_0)^2 + z_0^2}$.
Finally, the total magnetic field is simply the sum of the applied and induced field components $B(x,y) =  B_\textnormal{DC} + B_\textnormal{AC} + B_\textnormal{NP}(x,y)$.

Since all AC phenomena is restricted to frequencies well below the GHz range, a quasi-static approach for the transport in graphene is acceptable~\cite{liu2011graphene}.
Therefore, the effect of having magnetic fields with both AC and DC components acting on the sensor is the emergence of a Hall voltage with the same time dependence.
The amplitude of the in-phase AC Hall voltage can be calculated simply by taking differences of the Hall voltage obtained at the maximum and minimum values of the $B_\textnormal{AC}$ field, with or without the $B_\textnormal{DC}$ field and with or without a magnetic nanoparticle nearby.

We denote by $V_\textnormal{H}^{\alpha, \beta, \gamma}$ the Hall voltage calculated with $B =  \alpha B_\textnormal{DC} + \beta B_\textnormal{AC} + \gamma B_\textnormal{NP}$, where $\alpha = \{0,1\}$, $\beta = \{-1, +1\}$ and $\gamma = \{0, 1\}$.
Therefore, we express the amplitude of the AC component by $\left( V_\textnormal{H}^{\alpha, +1, \gamma} - V_\textnormal{H}^{\alpha, -1, \gamma} \right)$.
Consequently, the variation in the AC amplitude due to the application of a DC field is, thus,
\begin{equation}
\delta V_\textnormal{H}^{\gamma} \equiv \left(V_\textnormal{H}^{0, +1, \gamma} - V_\textnormal{H}^{0, -1, \gamma} \right) - \left(V_\textnormal{H}^{1, +1, \gamma} - V_\textnormal{H}^{1, -1, \gamma} \right) \, .
\label{eq:delta_VH}
\end{equation}
The change in the AC amplitude due to the DC field is affected by the presence of a nearby nanoparticle, which is controlled by parameter $\gamma$.
The device working conditions can be tailored as to create a substantial difference between the $\delta V_\textnormal{H}^\textnormal{1}$ (with NP) and $\delta V_\textnormal{H}^\textnormal{0}$ (no NP) levels, so that they can be distinguished experimentally.
In general, typical values of $V_\textnormal{H}^{\alpha, \beta, \gamma}$ range from $10^1$ to $10^3$ microvolts.
The amplitude of the Hall voltage AC component $\left( V_\textnormal{H}^{\alpha, +1, \gamma} - V_\textnormal{H}^{\alpha, -1, \gamma} \right)$ is, typically, of the order of $10^2\,\mu\textnormal{V}$, which should not pose a problem to state-of-the-art scientific measurement tools~\cite{hudson2014stochastic}.

\section*{Acknowledgement}
R. F. Neumann thanks Benoit Jouault (Universit\'e Montpellier 2/CNRS) for help with \texttt{FreeFem++}, Em\'{i}lio Vital Brazil and Rodrigo da Silva Ferreira (IBM Research) for fruitful discussions regarding the particle trajectory reconstruction algorithm. M. Engel thanks David W. Abraham and N. Wang (IBM Research) for discussion regarding device implementation.

\appendix

\section{Appendix}
We present in the Appendix some additional results to be used as a guideline for the choice of parameters.
Specifically, we discuss the choice of nanoparticle diameter and concentration, the choice of magnetic field components and present details of the numerical calculations.

\subsection{Choice of nanoparticle concentration}

\begin{figure}[ht]
\centering
\includegraphics[width=\columnwidth]{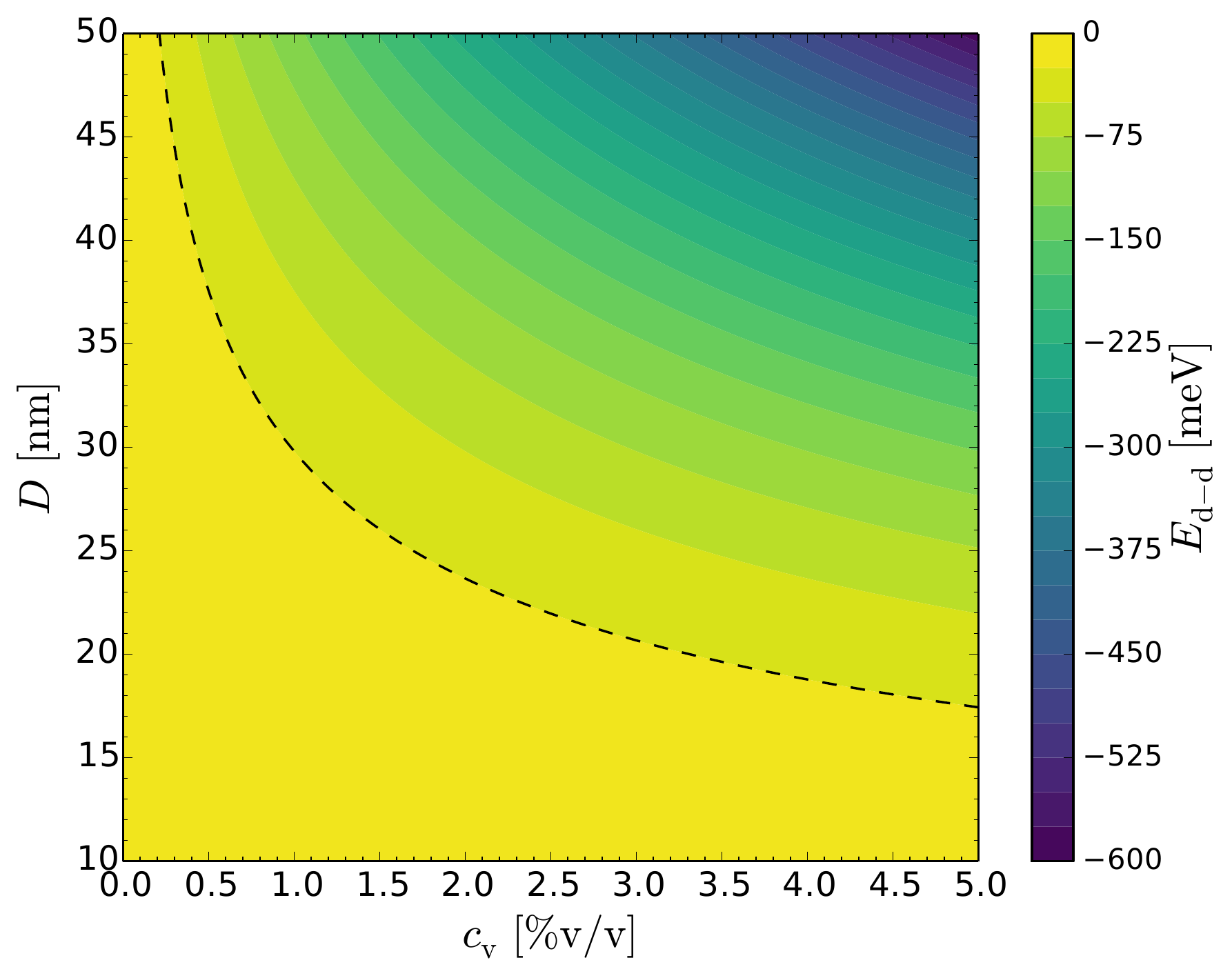}
\caption{Typical magnetic dipole-dipole energy as a function of nanoparticle concentration and diameter.
The dashed line represents $k_\textnormal{B} T$ at room temperature.}
\label{fig:EMAG_vs_kT}
\end{figure}

In order to help the reader select the nanoparticle volume/volume concentration $c_\textnormal{v}$ depending on the nanoparticle diameter $D$, Figure~\ref{fig:EMAG_vs_kT} shows the comparison between the typical magnetic dipole-dipole interaction energy $E_\textnormal{d-d} = -\frac{\mu_\textnormal{0}}{4 \pi} c_\textnormal{v} \left[ 2 M_\textnormal{S} \frac{\pi D^3}{6} \right]$ and the thermal energy $k_\textnormal{B} T$ at room temperature, for nanoparticles with diameter $D$ and saturation magnetization $M_\textnormal{S} = 380\,\mathrm{kA/m}$.
The formula for $E_\textnormal{d-d}$ was obtained by calculating the dipolar energy between neighbouring nanoparticles, assuming they are placed in a cubic lattice.
The dashed line represents $|E_\textnormal{d-d}| = k_\textnormal{B} T$ at room temperature.
Thermal agitation is expected to prevent agglomeration for $(c_\textnormal{v}, D)$ points well below the dashed line.

\subsection{Visualization of raw results}

\begin{figure}[ht]
\centering
\includegraphics[width=\columnwidth]{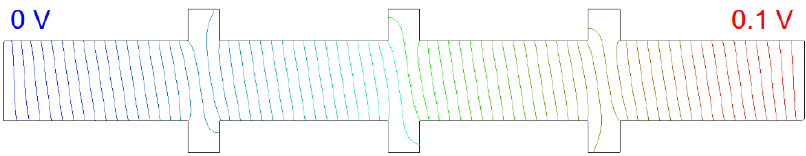}
\caption{Sketch of the numerical solution to $\phi(x,y)$ showing the equipotential lines that indicate the existence of a transverse Hall voltage $V_\textnormal{H}^{1, +1, 1}$ across the central measuring pad of a ``triple-cross'' geometry.}
\label{fig:V111}
\end{figure}

The numerical solution to Equation~\ref{eq:pde} provided by \texttt{FreeFem++} consists of a two-dimensional function $\phi(x,y)$ defined on a mesh within the sensor geometrical boundaries as in Figure~\ref{fig:sensor-sketch}.
In Figure~\ref{fig:V111} one can see the $\phi(x,y)$ used to calculate $V_\textnormal{H}^{1, +1, 1}$.
Table~\ref{tab:example} illustrates the typical raw values that enter Equation~\ref{eq:delta_VH}, as calculated by the current methodology.
\begin{table}[ht]
\centering
\begin{tabular}{l|cc}
\hline\hline
$[\mu V]$                           &   no NP       & with NP   \\ \hline
$V_\textnormal{H}^{0, +1, \gamma}$  &   74.39       &  75.84    \\
$V_\textnormal{H}^{0, -1, \gamma}$  &  -77.98       & -79.43    \\
$V_\textnormal{H}^{1, +1, \gamma}$  &   1595.91     &  1598.61  \\
$V_\textnormal{H}^{1, -1, \gamma}$  &   1444.10     &  1446.80  \\ \hline \hline
$\delta V_\textnormal{H}^{\gamma}$  &   0.56        &  3.46     \\
\end{tabular}
\caption{Individual Hall voltage values used to calculate $\delta V_\textnormal{H}^{0}$ and $\delta V_\textnormal{H}^{1}$ at the central measuring pad according to Equation~\ref{eq:delta_VH}.
This table corresponds to a $D = 20\,\textnormal{nm}$ nanoparticle at $z_\textnormal{0} = 100\,\textnormal{nm}$ above the origin and excited by $B_\textnormal{AC} = 5\,\textnormal{mT}$ and $B_\textnormal{DC} = 100\,\textnormal{mT}$ magnetic fields.}
\label{tab:example}
\end{table}

\subsection{Choice of magnetic field components}

\begin{figure}[ht]
\centering
\begin{subfigure}[h]{0.4\textwidth}
    \caption{\hfill~}
    \includegraphics[width=\columnwidth]{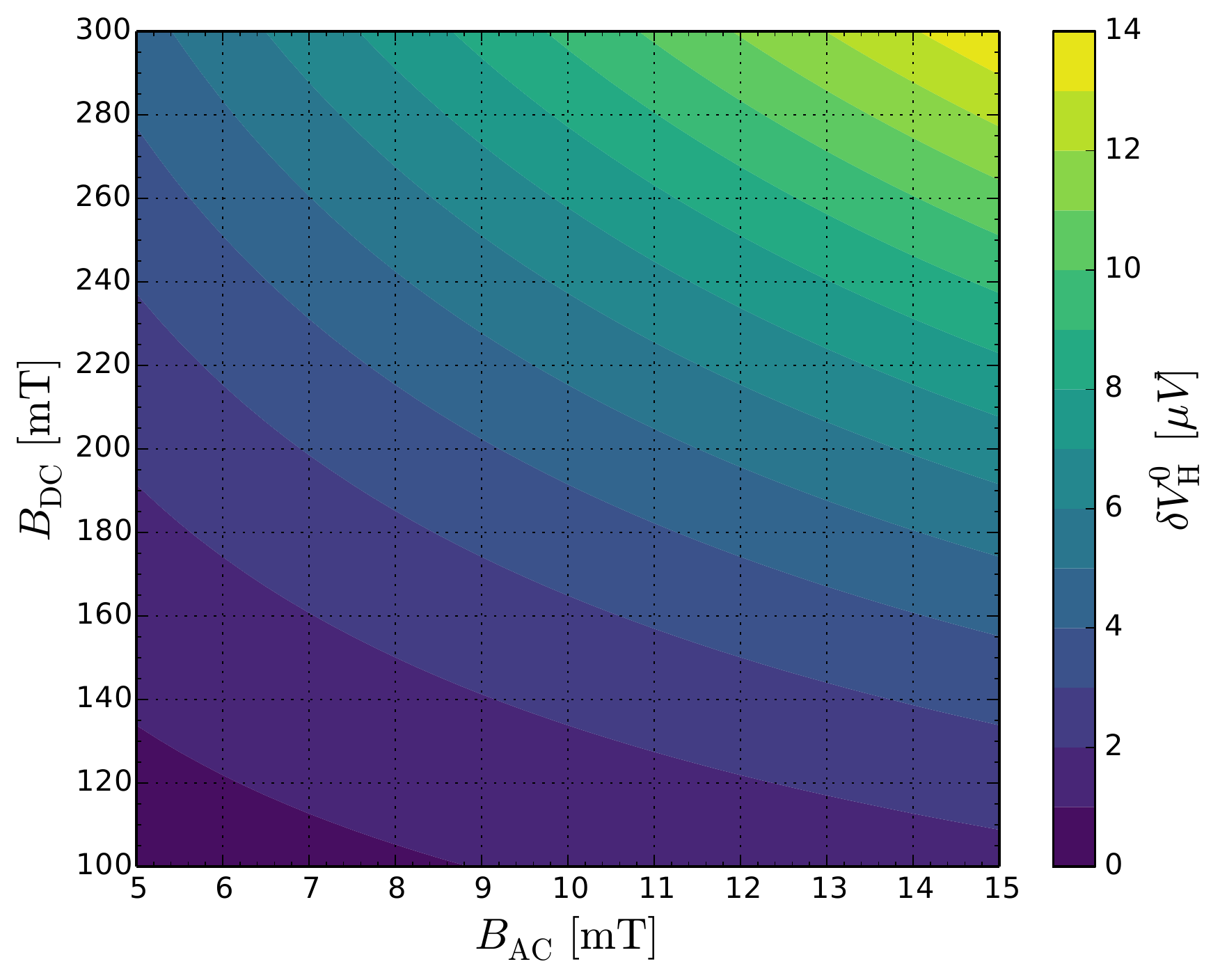}
    \label{fig:Baseline-Field}
\end{subfigure}
\begin{subfigure}[h]{0.4\textwidth}
    \caption{\hfill~}
    \includegraphics[width=\columnwidth]{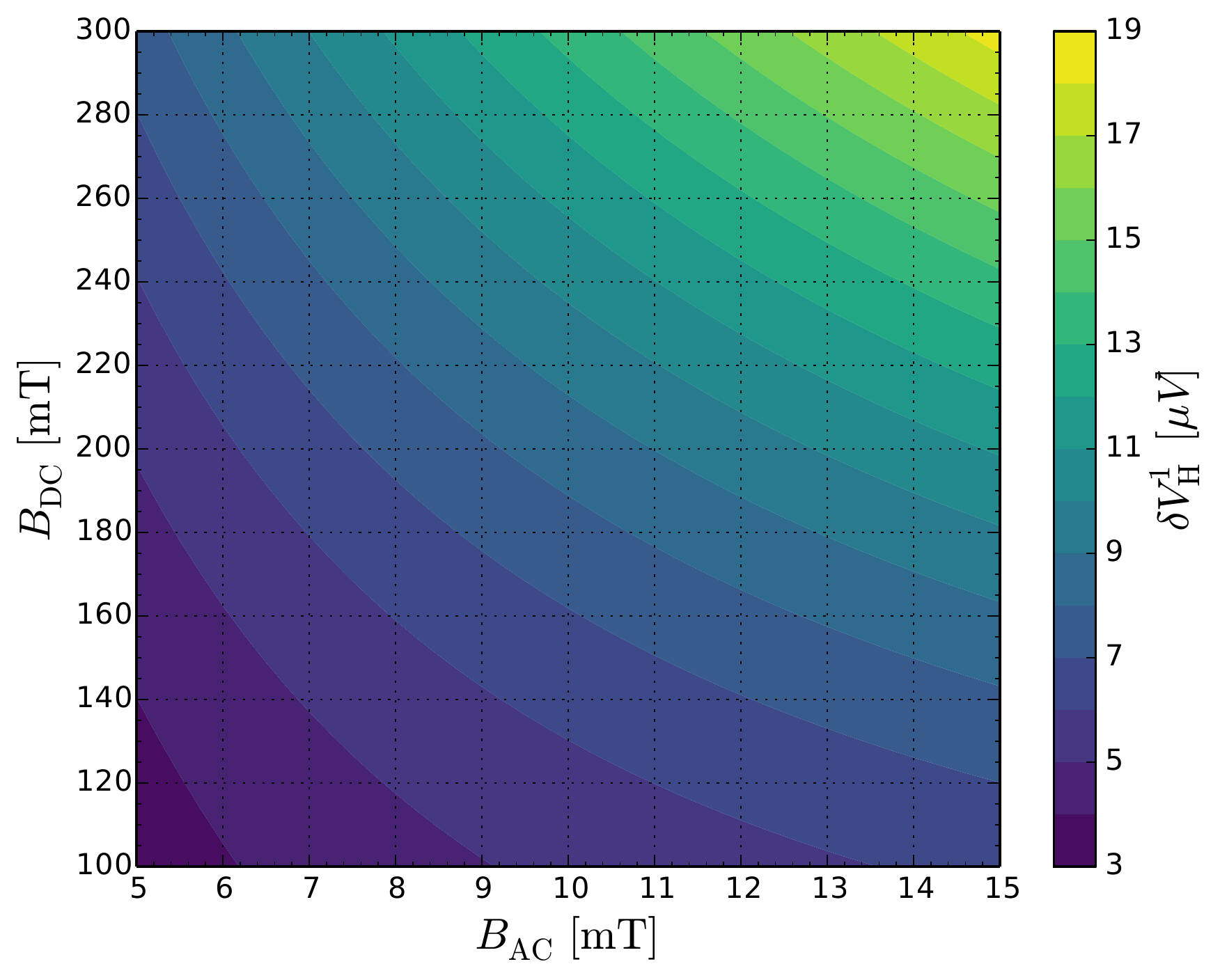}
    \label{fig:VH-Bfield}
\end{subfigure}
\begin{subfigure}[h]{0.4\textwidth}
    \caption{\hfill~}
    \includegraphics[width=\columnwidth]{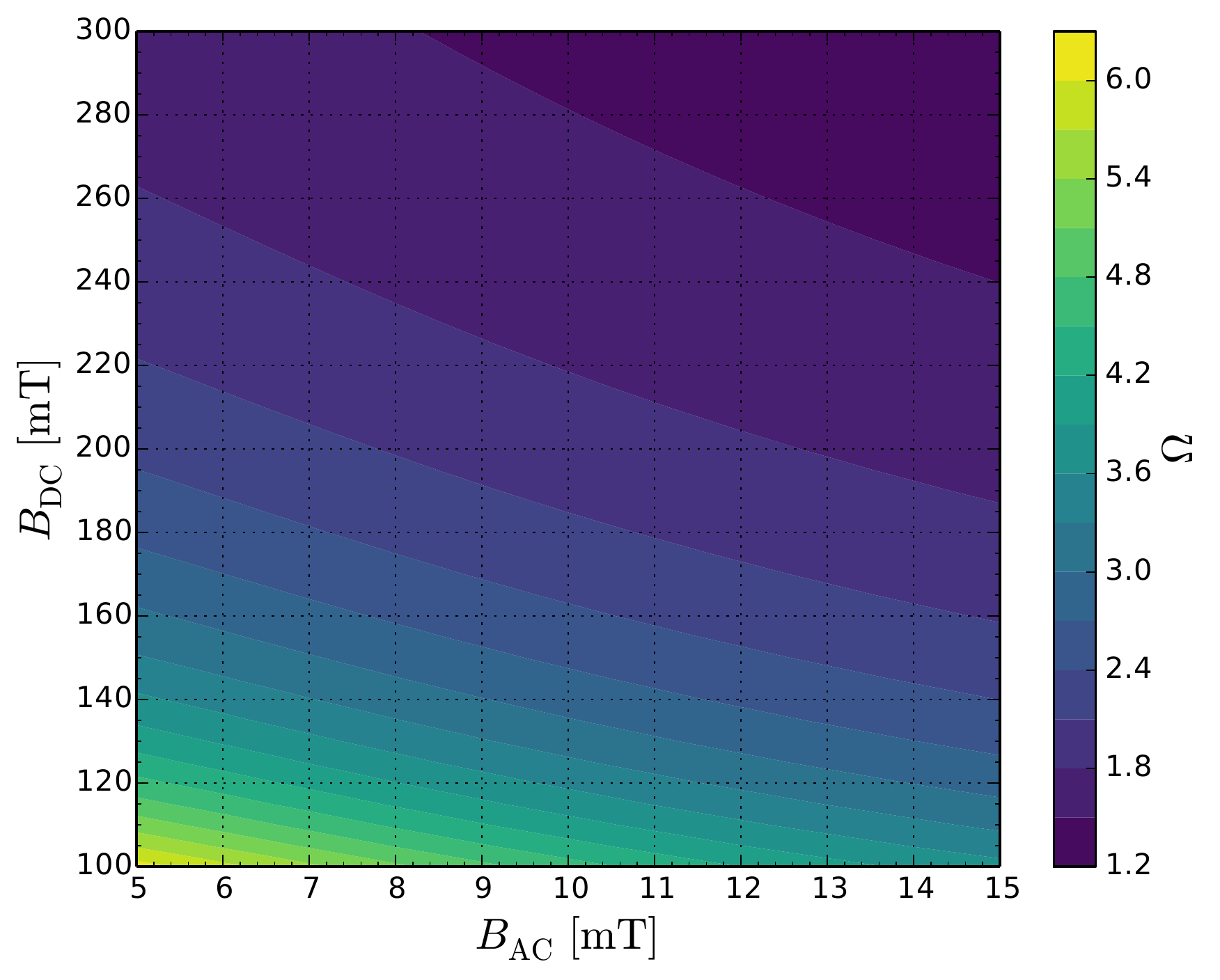}
    \label{fig:RelChange-Bfield}
\end{subfigure}
\caption{Variation of the Hall voltage AC amplitude $\delta V_\textnormal{H}^{\gamma}$ as a function of $B_\textnormal{DC}$ and $B_\textnormal{AC}$, for when there is (\subref{fig:Baseline-Field}) no nanoparticle nearby and  (\subref{fig:VH-Bfield}) a $D = 20\,\textnormal{nm}$ nanoparticle placed $z_\textnormal{0} = 100\,\textnormal{nm}$ from the origin. Panel (\subref{fig:RelChange-Bfield}) represents the ratio $\Omega_\textnormal{C} = \left( \delta V_\textnormal{H}^{1} / \delta V_\textnormal{H}^{0} \right)\!\big|_C$ evaluated at the central measuring pad.}
\label{fig:Bfields}
\end{figure}

For the choice of the AC and DC components of the applied magnetic field, several combinations were evaluated in terms of the relative change $\Omega_m = \left( \delta V_\textnormal{H}^{1} / \delta V_\textnormal{H}^{0} \right)\!\big|_m$ in the Hall voltage AC amplitude they induce.
In Figure~\ref{fig:Bfields} one finds the results for the detection of a $D = 20\,\textnormal{nm}$ nanoparticle $z_\textnormal{0} = 100\,\textnormal{nm}$ away from the origin.
The detection signal $\delta V_\textnormal{H}^{1}$ in Figure~\ref{fig:Bfields}(\subref{fig:VH-Bfield}) changes drastically with respect to the baseline level $\delta V_\textnormal{H}^{0}$ in Figure~\ref{fig:Bfields}(\subref{fig:Baseline-Field}), specially for low $B_\textnormal{AC}$ fields.
As a result, the relative change in the signal that allows for the detection of an adjacent nanoparticle attains higher values on the lower left corner of Figure~\ref{fig:Bfields}(\subref{fig:RelChange-Bfield}).
The 6-fold increase in the signal for $B_\textnormal{AC} = 5\,\textnormal{mT}$ and $B_\textnormal{DC} = 100\,\textnormal{mT}$ is large enough for most experimental implementations and, therefore, those field values were chosen. For larger nanoparticles, even higher $\Omega$ values are obtained.

\subsection{Choice of nanoparticle size}

\begin{figure}[ht]
\centering
\begin{subfigure}[h]{0.4\textwidth}
    \caption{\hfill~}
    \includegraphics[width=\columnwidth]{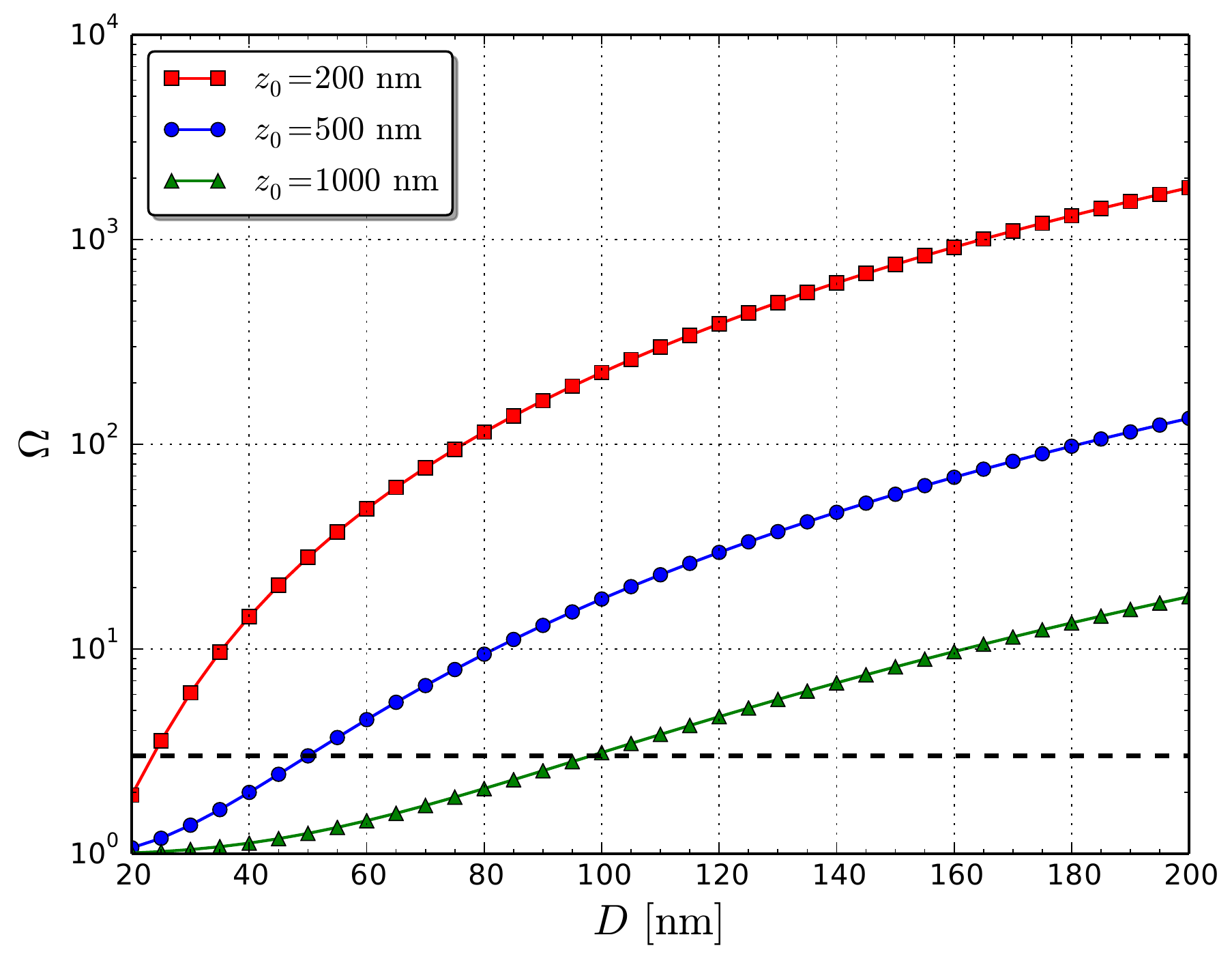}
    \label{fig:Omega-Dz_D}
\end{subfigure}
\begin{subfigure}[h]{0.4\textwidth}
    \caption{\hfill~}
    \includegraphics[width=\columnwidth]{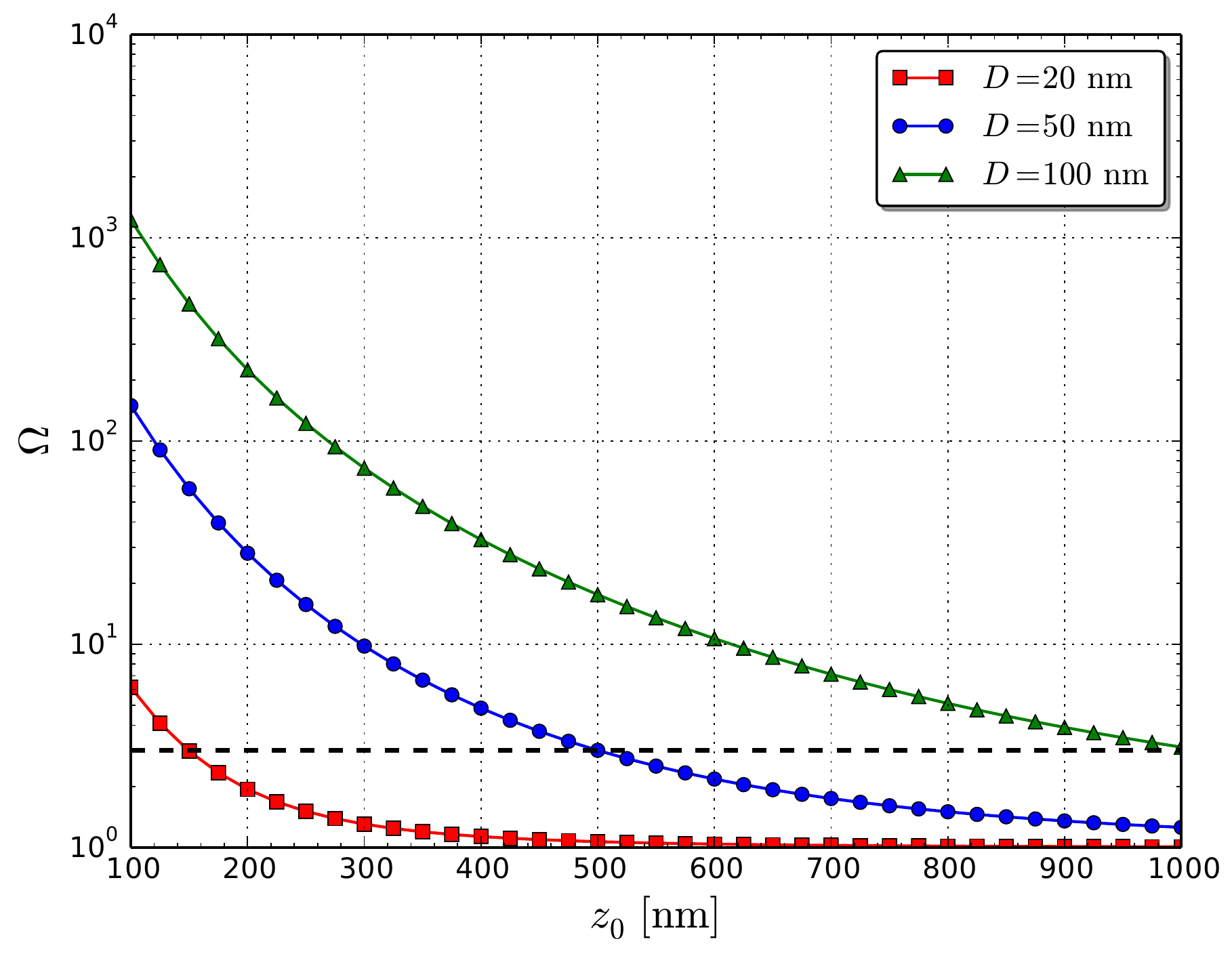}
    \label{fig:Omega-Dz_z}
\end{subfigure}
\caption{Selected subplots showing the variation of $\Omega$ (\subref{fig:Omega-Dz_D}) along $D$ for selected $z_\textnormal{0}$ and (\subref{fig:Omega-Dz_z}) along $z_\textnormal{0}$ for selected $D$.
The dashed line represents $\Omega = 3$.}
\label{fig:Dz}
\end{figure}

As presented in Figure~\ref{fig:Omega-Dz}, the nanoparticle diameter $D$ must be chosen in accordance with the channel height, for it limits the maximum vertical distance $z_\textnormal{0}$ at which the nanoparticle is detectable.
In order to facilitate the selection of diameters, Figures~\ref{fig:Dz}(\subref{fig:Omega-Dz_D}) and~\ref{fig:Dz}(\subref{fig:Omega-Dz_z}), respectively, show the dependency of $\Omega$ on $D$ for selected $z_\textnormal{0}$ and on $z_\textnormal{0}$ for selected $D$.
For a target sensitivity of $\Omega = 3$, represented by a dashed line, one clearly concludes that $D = 50\,\textnormal{nm}$ and $100\,\textnormal{nm}$ nanoparticles are required for channels as high as $500\,\textnormal{nm}$ and $1000\,\textnormal{nm}$ respectively.

\bibliographystyle{unsrt}
\bibliography{article}
\end{document}